\pdfoutput=1
\documentclass[showpacs,prb,twocolumn]{revtex4}
\usepackage{amsmath}
\usepackage{amssymb}
\usepackage{color}
\usepackage{graphics}
\usepackage{graphicx}
\usepackage{dcolumn}

\begin{document}

\title {Resonant plasmonic effects in periodic graphene antidot arrays}

\author{ A.~Yu.~Nikitin$^{1}$}
\email{alexeynik@rambler.ru}
\author{ F. Guinea$^2$}
\author{ L.~Martin-Moreno$^1$}
\email{lmm@unizar.es}
 \affiliation{$^1$ Instituto de Ciencia de Materiales de Arag\'{o}n and Departamento de F\'{i}sica de la Materia Condensada,
CSIC-Universidad de Zaragoza, E-50009, Zaragoza, Spain \\
$^2$ Instituto de Ciencia de Materiales de Madrid, CSIC, Cantoblanco, E-28049 Madrid, Spain}

\begin{abstract}

We show that a graphene sheet perforated with micro- or nano-size antidots have prominent absorption resonances in the microwave and terahertz regions. These resonances correspond to surface plasmons of a continuous sheet ``perturbed'' by a lattice. They are excited in different diffraction orders, in contrast to cavity surface plasmon modes existing in disconnected graphene structures. The resonant absorption by the antidot array can essentially exceed the absorption by a continuous graphene sheet, even for high antidot diameter-to-period aspect ratios. Surface plasmon-enhanced absorption and suppressed transmission is more efficient for higher relaxation times of the charge carriers.

\end{abstract}

 \maketitle

Graphene now is widely known not only for its amazing transport properties,\cite{Review09} but also for a significant potential in photonics.\cite{graphphot_natphot10} Most applications require that the electromagnetic (EM) field is first absorbed by graphene. In a large part of the EM spectrum the two-dimensional conductance of one monolayer $\sigma$ is small, so that $\alpha=2\pi\sigma/c\ll1$. Then the absorption coefficient $A$ for a free-standing graphene can be approximately written at normal incidence as $A\simeq2 \mathrm{Re}(\alpha)$. In the optical regime graphene behaves as an absorbing dielectric (its effective refractive index is a purely imaginary number): $\alpha_{\textrm{opt}}=\pi\alpha_0/2$, with $\alpha_0=1/137$ being the fine-structure constant. This leads to the classical result $A_{\textrm{opt}}=\pi\alpha_0\simeq2.3\%$. In both terahertz (THz) and microwave frequency ranges the situation is rather different. The conductivity is contributed predominantly by the intra-band electronic transitions, so that graphene behaves like a two-dimensional metal. For low temperatures its conductance can be approximated by a Drude-Lorentz model\cite{Wunsch06,Hwang07,Falkovsky08} $\alpha_{\textrm{THz}}=2i\alpha_0|\mu|/\hbar(\omega+i\tau^{-1})$ depending upon the chemical potential $\mu$, frequency $\omega$ and relaxation time $\tau$. In this case for $\omega\tau\ll 1$ the absorption coefficient reads $A_{\textrm{THz}}\simeq 4\alpha_0|\mu|/(\hbar\tau\omega^2)$ and can become of order of tenths of a percent. Obviously, it is challenging to use graphene as an efficient absorber.

Fortunately, one of the promising points of graphene is that it supports surface waves [graphene surface plasmons (GSPs)].\cite{Shung86,Campagnoli89,Vafek06,Hansonw08,IRSPP09,Engheta,PlasmonicsNature11,StauberPRL11,NikitinPRBR12,deAbajoPRL12,ringsAPL12,KoppensArXive12,BasovArXive12} GSPs allow for a strong subwavelength confinement of the EM fields and thus for a much higher absorption. Previously it has been demonstrated that periodic arrays of graphene ribbons,\cite{PlasmonicsNature11,NikitinPRBR12} discs\cite{deAbajoPRL12} and rings\cite{ringsAPL12} can resonantly absorb THz radiation due to excitation of GSP cavity modes. However, this kind of structures is not ``electrically-continuous'' preventing therefore a direct current passing through the samples. An alternative solution can be a graphene periodic antidot array (GPAA),\cite{PedersenPRL08,KimNL10,nanomesh10,AkhavanACSNano10,KraussNL10,BegliarbekovNL11} which presents both continuity (to provide the electric transport) and periodicity (to transfer the momentum to GSPs from the incident wave). The antidots in a graphene sheet can be considered as regions of zero conductivity due to either (i) actual holes in graphene, (ii) the influence of an external voltage or (iii) appropriate chemical doping.

GPAAs could open interesting possibilities for electromagnetic control via GSP-assisted effects. This can be an important point for designing photo-electric devices, e.g. tunable absorbers that can act at the same time as conducting electrodes.

\begin{figure}[thb!]
\includegraphics[width=7cm]{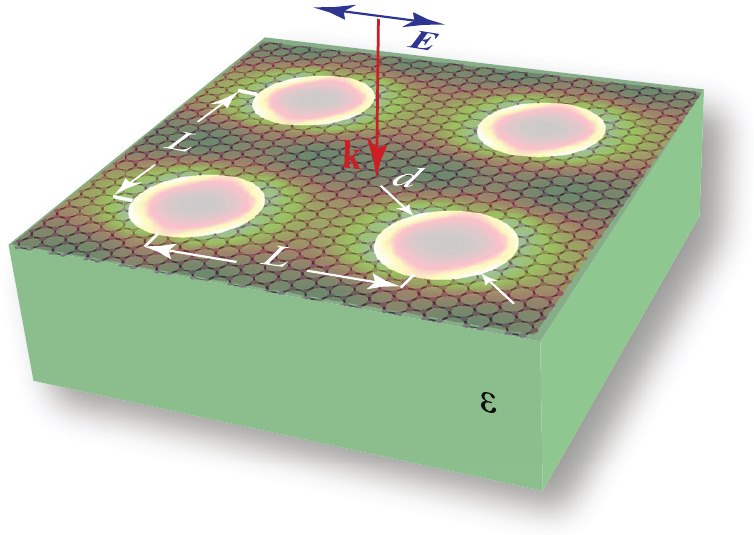}\\
\caption{(Color online)  The schematic of the studied system: a plane electromagnetic wave impinging onto a periodic antidot array in a graphene monolayer placed onto a dielectric substrate.}\label{geom}
\end{figure}

In this Letter we report on GSP-enhanced absorption and suppressed transmission in continuous GPAAs with circular antidots. We characterize resonances, presenting various dependencies of absorption $A$, transmission $T$, and reflection $R$ coefficients upon several parameters: period $L$, antidot diameter $d$ and relaxation time $\tau$.  We consider a square GPAA placed on the substrate with dielectric permittivity $\varepsilon$. The system is illuminated by a normal-incident monochromatic plane wave with wavevector $k$ from the vacuum half-space. The electric field of the incident wave is parallel to one of the Bragg's vectors of the lattice. The scheme of the structure is shown in Fig.~\ref{geom}. Graphene is modeled by a conductivity $\sigma$, computed within the random-phase approximation,\cite{Wunsch06,Hwang07,Falkovsky08} which is valid provided that the mean free path is shorter than the superlattice scale. In this work we assume room temperature, $T=300$~K.

\begin{figure}[thb!]
\includegraphics[width=8cm]{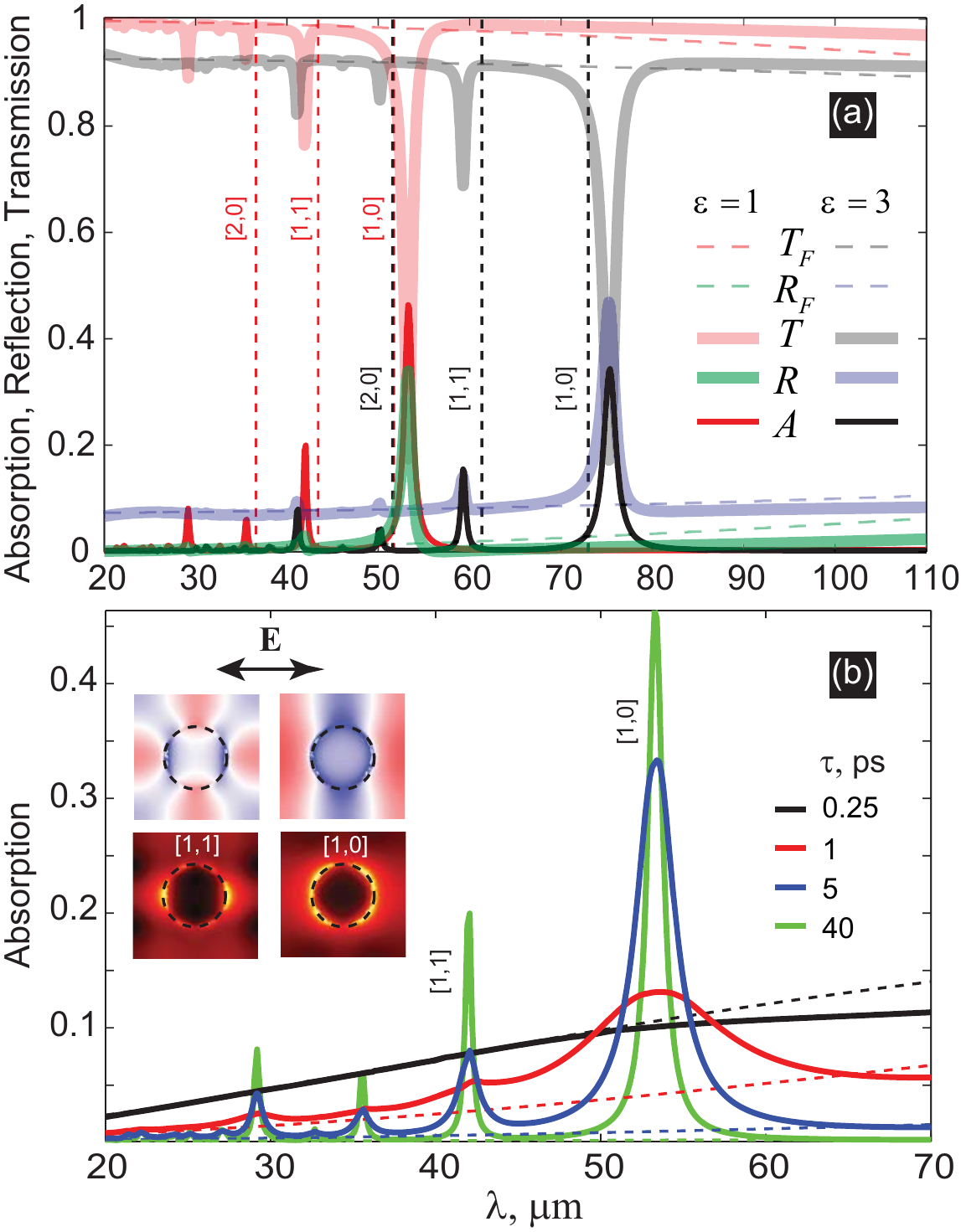}\\
\caption{(Color online)  The spectra of absorption, reflection and transmission coefficients for square GPAAs. (a) $A$, $R$, $T$ for the suspended GPAA and GPAA on the substrate with $\varepsilon=3$ and $\tau=40$~ps. $R_F$ and $T_F$ corresponding to the continuous graphene sheet (Fresnel coefficients) are shown by the discontinuous curves.  (b) Absorption spectra for GPAAs at different $\tau$ for $\varepsilon=1$. The spectra for continuous graphene monolayers are shown by the dashed lines. The insets to (b) show the colorplots for the spatial distribution of the electric field: modulus (below) and the real part of the projection on the direction of the incident wave electric field shown by an arrow (above). In all panels $\mu=0.2 eV$, $L=2d=5 \mu $m.}\label{spectra}
\end{figure}

An example of the wavelength spectra for $A$,  $T$ and $R$ is shown in Fig.~\ref{spectra} (a) corresponding to $L=2d=5 \mu $m. The simulations have been conducted by using the finite elements method (FEM).\cite{COMSOL} We choose this value of $L$ for demonstration purposes, in order to have prominent resonances in the window $20-100 \mu$m; the dependence with $L$ will be addressed later on.
Both the case of free-standing GPAA and GPAA on the substrate are shown. While absorption and reflection coefficients have a set of resonance maxima, the transmission spectra possess pronounced minima. The positions of the resonances correspond approximately to the condition of GSP excitation in different diffraction orders, i.e. $G\sqrt{n^2+m^2} = k_p$, where $G=2\pi/L$ is the shortest Bragg's vector, and $k_p$ is the GSP wavevector. For convenience, these resonances will be further labeled as $[n,m]$. In suspended GPAAs with sufficiently small antidots a good approximation for $k_p$ can be $k_p \simeq k\sqrt{1-\bar{\alpha}^{-2}}$, with $\bar{\alpha} = \alpha\cdot[1-\frac{\pi}{4}(\frac{d}{L})^2]$ being the average value of the normalized graphene conductance in PGAA. The characterization of the resonance positions for large antidots will be presented below. For graphene in a nonsymmetric dielectric environment the dispersion relation reads $\frac{\varepsilon_1}{q_{z1}}+\frac{\varepsilon_2}{q_{z2}}+2\alpha=0$, where $q_{z1,2}=\sqrt{\varepsilon_{1,2}-k^2_p/k^2}$ and $\varepsilon_{1,2}$ are the dielectric constants of the surrounding media.\cite{Hansonw08} This equation cannot be resolved in algebraic functions and therefore is treated numerically. The positions of the three highest resonance wavelengths estimated from the above ``undressed'' dispersion law are shown by the discontinuous vertical lines in Fig.~\ref{spectra} (a). The minimum/maximum corresponding to the longest wavelength resonance $[1,0]$  is redshifted with respect to the unperturbed resonance wavelength, while the next two (lower-wavelength) ones, i.e. $[1,1]$ and $[2,0]$ are blue-shifted. In a symmetric configuration the maximal limiting value for $A$ is $1/2$, but it can be increased, for instance, placing a metal layer below the GPAA.\cite{NikitinPRBR12,deAbajoPRL12}

In order to better visualize the resonance picture, the inset to Fig.~\ref{spectra}~(b) shows the spatial field distribution  for both electric field modulus and the real part of the electric field component along the polarization of the incident wave. The colorplots are taken at the wavelengths corresponding to $[1,0]$ and $[1,1]$ resonances at $\lambda = 53.26 ~\mu $m and $\lambda = 42.06 ~\mu $m, respectively. Let us point out that the plasmonic wavelengths $\lambda_p$ corresponding to a continuous monolayer for these values of $\lambda$ are $\lambda_p = 6.61 ~\mu $m and $\lambda_p = 4.12 ~\mu $m.

Notice that GSP resonances in PGAA are reminiscent to short-range surface plasmon resonances studied in hole arrays in ultra-thin metallic films.\cite{RodrigoOL09} However, in case of metallic structures, absorption and reflection peaks (transmission dips) appear in the optical region.

Owing to much higher values of $\tau$,\cite{HwangPRL07,BolotinSSC08} suspended samples can present a special interest. Nowadays, placing graphene on a substrate severely reduces the mobility. For this reason, in this Letter we mainly concentrate on the free-standing GPAA. However, for academic purposes, in order to see other effects of the substrate not related to the change of mobility, in Fig.~\ref{spectra} (a) we compare the spectra for symmetric and nonsymmetric configurations at the same value of $\tau$.   From this comparison it follows that the spectra  are similar, and the only significant difference between them consists in a displacement of the resonance for GPAA to longer wavelength compared to the free-standing case.  This is related to the higher values of the GSP wavevector for a nonsymmetric dielectric environment. Let us turn to the behavior of the absorption coefficient with respect to the change of $\tau$.

\begin{figure}[thb!]
\includegraphics[width=8cm]{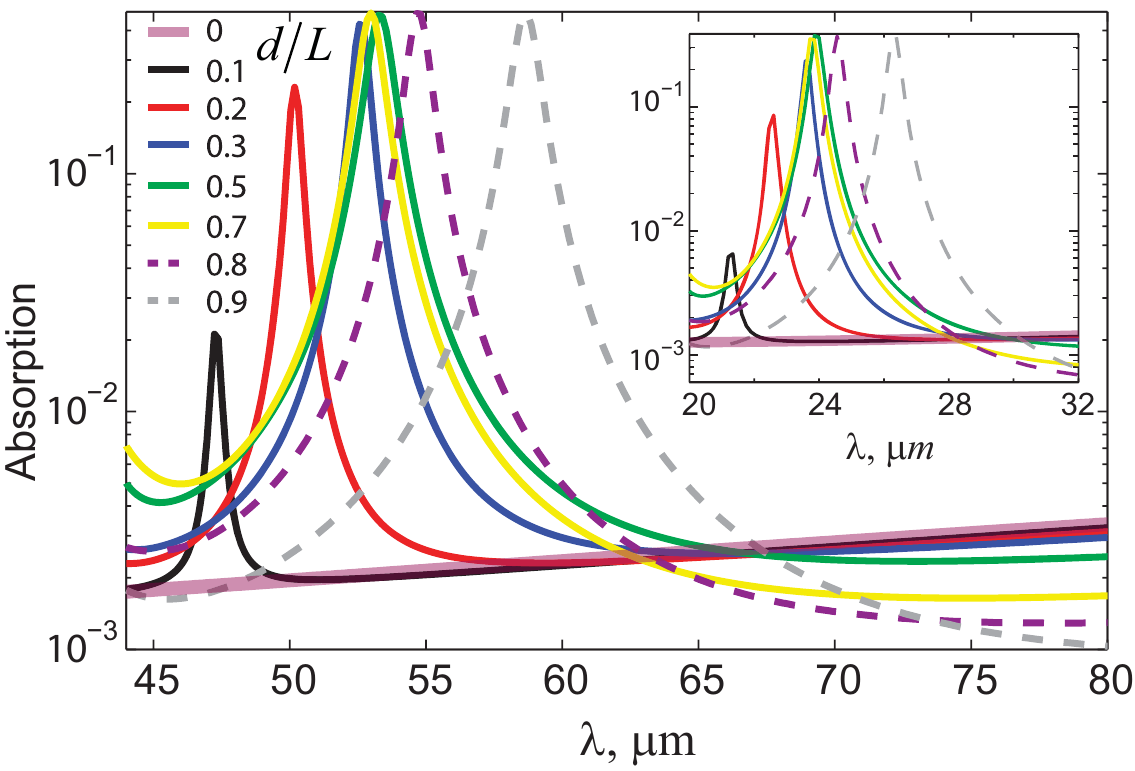}\\
\caption{(Color online)  (c) Absorption spectra for GPAAs at different antidot sizes $d$. The main figure is for $L=5 \mu $m, while the inset is for $L=1 \mu $m. The curve corresponding to $d/L=0$ renders $A$ for the continuous monolayer. $\varepsilon=3$, $\tau=40$~ps, $\mu=0.2 eV$.}\label{spectrad}
\end{figure}

In Fig.~\ref{spectra} (b) the absorption spectra for different relaxation rates $\tau$ are compared. It is seen that higher $\tau$ provide higher visibility of the resonances, especially for lower wavelengths. An important message here is that the GSP-induced absorption in GPAAs can not only be higher than the absorption corresponding to lower $\tau$, but can also largely exceed the absorption in the continuous graphene monolayer. Since  resonances for different $\tau$ show the same behavior in function of geometric parameters of GPAAs, we will further concentrate on the case of the maximal predicted value for $\tau$ in continuous films.\cite{HwangPRL07}

The resonances are very sensitive to the geometrical aspect ratio $d/L$.
In Fig.~\ref{spectrad} the evolution of the longest-wavelength resonance $[1,0]$ with respect to the antidot diameter change is illustrated for $L=5~ \mu $m. When the antidot size increases, the resonance suffers a redshift and the intensity of the absorption peak increases.  However, in the interval of $d/L = (0.3,0.8)$ the resonance position is quite stable. Further increasing $d/L$ up to the values $d/L\sim1$ leads to the fast displacement of the peak. A point to note is that even for large $d/L$, for which the graphene area is smaller than the one of the antidot one, the maximal value of the resonance peak remains hight. The same behavior takes place for other values of the period. As an example, the case of $L=1~ \mu $m is shown in the inset to Fig.~\ref{spectrad}. Notice that in contrast to the arrays of graphene ribbons or discs, the resonance position for GPAA at a fixed value of $d$ is strongly dependent upon the period due to Bragg's origin of the GSP resonance.

\begin{figure}[thb!]
\includegraphics[width=8cm]{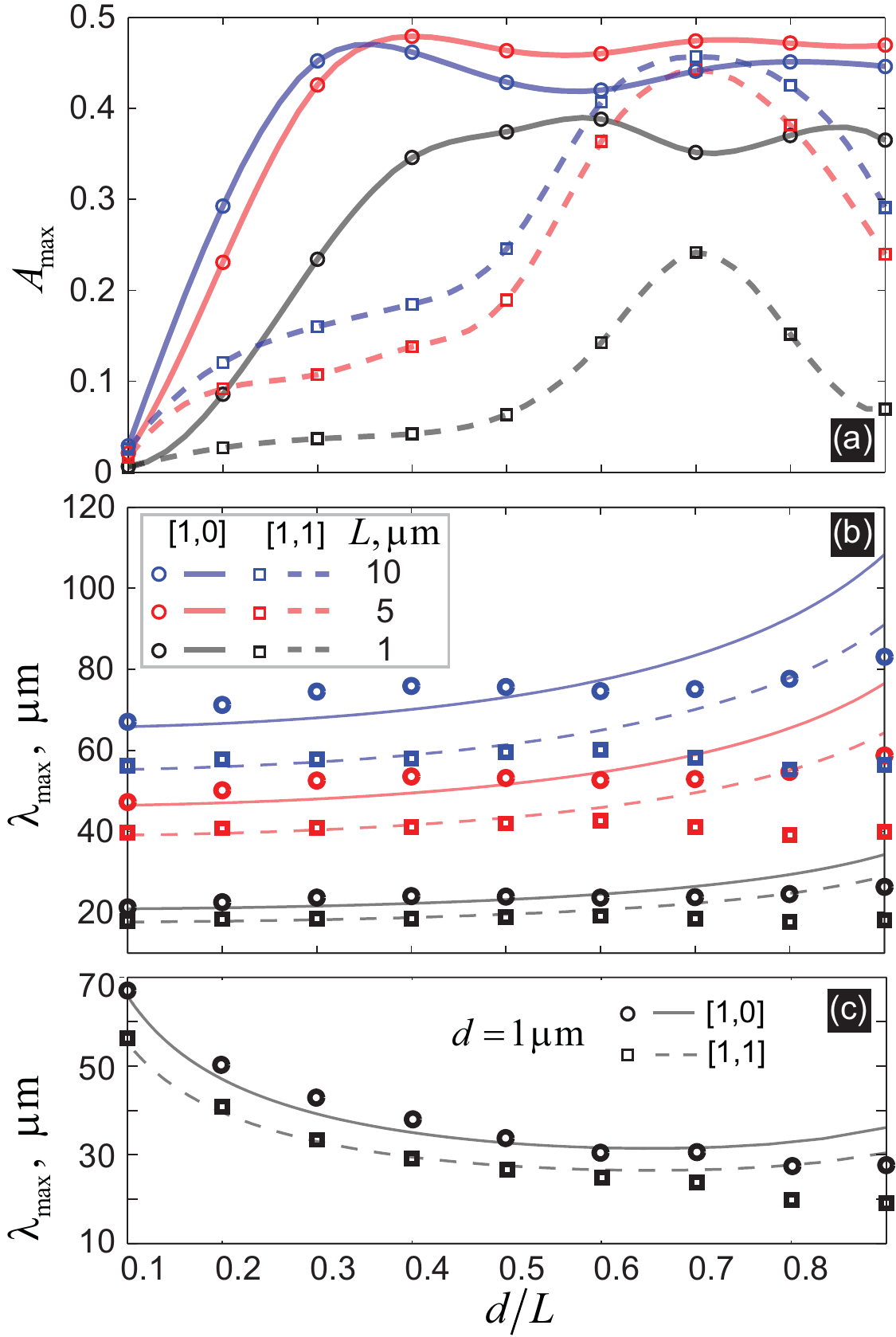}\\
\caption{(Color online)  Characterization of the resonances in GPAAs with respect to the aspect ratio $d/L$. (a) Maximal value of absorption $A_{\mathrm{max}}$ as a function of $d/L$ at a fixed $L$ for the resonances $[1,0]$ and $[1,1]$. (b) The positions of the maxima for $[1,0]$ and $[1,1]$ resonances at different $d/L$ for a fixed $L$. (c) The positions of the maxima for $[1,0]$ and $[1,1]$ resonances at different $d/L$ for a fixed $d$. In (b), (c) continuous lines represent the estimated position of the resonance, while the symbols correspond to the full calculations. In all panels $\varepsilon=1$, $\mu=0.2 ~eV$ and $\tau=40$~ps.}\label{position}
\end{figure}

In order to gain a further insight into the properties of GSP resonances in GPAAs, let us analyze in more detail the dependencies of the absorption maxima $A_{\mathrm{max}}$ and resonance wavelength $\lambda_{\mathrm{max}}$  upon $d/L$. We consider GPAAs with the $L$ values down to $1 \mu$m and values of $d$ down to $100$nm. In Fig.~\ref{position}~(a),(b) $A_{\mathrm{max}}$ and $\lambda_{\mathrm{max}}$ both for $[1,0]$ and $[1,1]$ resonances are shown as a function of $d$, at several fixed periods. For the $[1,0]$ resonance, $A_{\mathrm{max}}$ quickly reaches high values, after which it saturates, remaining practically constant for all periods. In contrast, $A_{\mathrm{max}}$ for the $[1,1]$  resonance has an optimum close to $d/L=0.7$, with this value being independent upon $L$.

The dependency of $\lambda_{\mathrm{max}}$ upon $d$ and $L$ changes in different regions of $d/L$. When $d/L\lesssim0.1$, the dispersion relation of the GP and therefore $\lambda_{\mathrm{max}}$,  is governed by the average value of $\alpha$, see the coincidence of the points and the curves in the region of small $d/L$ in  Fig.~\ref{position}~(b,c). However, for larger $d/L$ there is a modification of the GSP dispersion branches. The degenerated resonance diffraction orders become strongly coupled (directly and also due to multiple scattering processes through both evanescent and radiative diffraction orders), leading to an increase of the bandgap width and a shift of the bandgap center. In other words, the dependence of $\lambda_{\mathrm{max}}$ deviates from the estimation based on the average conductance, see Fig.~\ref{position}~(b,c).

To summarize, in this Letter we have shown that graphene periodic antidot arrays (GPAAs) can provide a strong electromagnetic response in both microwave and THz regions, where GSPs are excited. The absorption peaks and transmission dips in the antidot structure are more pronounced for less absorbing graphene samples, characterized by higher relaxation rates of charge carriers. The system discussed here can be realized in CVD graphene, where large samples can be routinely fabricated.\cite{BaeNatNan10} The resonances in antidot arrays are due to excitation of dressed GSPs under the Bragg's condition. This makes the difference with electrically-isolated graphene periodic structures (stripes, discs, etc.), where the resonances are due to cavity GSP-modes. The continuous graphene structures possess thus a supplementary tunability parameter $L$. Additionally, GPAAs can have applications for devices, in which the same graphene layer can be both an electric conductor and radiation absorber.

The authors are grateful to M. L. Nesterov for
helpful discussions regarding numeric simulations, and to Institute for Biocomputation and Physics of Complex Systems (BIFI) of Zaragoza for computational reasons. We acknowledge support from the Spanish MECD under Contract No. MAT2009-06609-C02, FIS2008-00124, Consolider CSD2007-00010, and Consolider Project ``Nanolight.es''. A.Y.N. acknowledges the Juan de la Cierva Grant No. JCI-2008-3123.


\begin{thebibliography}{99}
\bibitem{COMSOL} A commercial software \textsc{comsol} has been used.
\end{thebibliography}
\end{document}